\documentclass[aps,twocolumn,pra,showpacs,amsmath,nofootinbib]{revtex4}
\usepackage{graphicx}% Include figure files
\usepackage{dcolumn}% Align table columns on decimal point
\usepackage{bm}% bold math
\usepackage{epstopdf}
\usepackage{multirow}
\newcommand{\beq}{\begin{equation}}
\newcommand{\eeq}{\end{equation}}
\newcommand{\bea}{\begin{eqnarray}}
\newcommand{\eea}{\end{eqnarray}}
\newcommand{\eps}{\varepsilon}
\newcommand{\frc}[2]{\raisebox{1pt}{$#1$}\big/\raisebox{-1pt}{$#2$}}

\begin{document}

\author{S. Kamerdzhiev\footnote{kamerdzhievsp@nrcki.ru}}
\affiliation{National Research Center Kurchatov Institute, Moscow,
123182 Russia.}

\author{M. Shitov}
\affiliation{National Research Center Kurchatov Institute, Moscow,
123182 Russia.}

\title{On microscopic theory of pygmy- and giant resonances }

%\pacs{21.60.−n, 74.20.−z}

\begin{abstract}
 The Green function formalism with a consistent account for  phonon coupling (PC), based on the self-consistent theory of finite Fermi-systems,
 is applied for pygmy- and giant multipole resonances in magic nuclei  with the aim to consider  particle-hole (ph) and complex 1p1h$\otimes$phonon configurations. A new equation for the effective field, which describes nuclear polarizability, is obtained. It contains  new PC contributions    %effective field  have been  obtained
  which are of interest in the energy region under consideration. They are due to:
  i)the tadpole effect  in the standard  ph-propagator, ii)two new induced interactions (caused by the exchange of  ph-phonon)  in the second
 ph-channel and in the particle-particle channels, and  iii)  the first and second variations of the effective interaction in the phonon field. The general expressions for energies and  probabilities of transitions between the ground and excited states  are obtained.  The qualitative analysis and discussion  of the new terms are performed.

\end{abstract}

\maketitle

\section{introduction}
The modern microscopic theory of pygmy- and  giant multipole resonances (for simplicity, hereafter  PDR and GMR) is mainly at the level of accounting  for:
i)  self-consistency between the mean field and effective interaction and 
ii) the low-lying phonon coupling (PC) \cite{Paar,Bracco,kaevYadFiz2019}. First of all, accounting for PC explained the third  integral GMR characteristic, i.e. the width \cite{revKST, Paar,ave2011}. The self-consistency approach turned out very useful for astrophysical applications even at QRPA level  due to its great predictive power \cite{Goriely}. Also, PC inclusion  resulted in   considerable improvement in explanations of  GMRs  properties  and appropriate applications \cite{Paar, Bracco, kaevYadFiz2019, YadFiz2011}, although some   problems
are encountered even  for magic nuclei. For example, the ones connected with  the  Skyrme parameters \cite{Tselyev2016} and spin-spin forces \cite{Lyutor2018} universal for all nuclei.

     As far as PDR is concerned, the situation  is more complex and requires more detailed consideration. Firstly, it is due to  existence of many new and delicate physical effects in this energy region, like irrotational and  vortical kind of   motion (the toroidal dipole resonance  which is  present in all the nuclei, while PDR  manifests itself in the nuclei with N$ > $Z \cite{Nester, Bracco,Paar}), and the  upbend phenomenon \cite{Larsen} at 1-3 MeV energy region. Besides, there are new  experimental possibilities \cite{Cosel,Larsen, Bracco}, for example, polarized proton inelastic scattering at very forward angles \cite{Cosel}. 
Secondly, as it turned out, it is impossible to explain completely the observed PDR fine structure within the latest self-consistent approach with Skyrme functional and accounting  for PC \cite{Lyutor2018} in the nucleus $^{208}$Pb, which was always a polygon for theory and experiment in PDR and GMR fields.  All these facts  
 can be a real reason  for   an additional analysis of PDR and GMR microscopic theory.
 
     In the field of  the self-consistent description  of characteristics of ground and low-lying collective states,  large work and a lot of  succesful calculations have been performed by the Kurchatov Institute group, see  reviews \cite{SapTol2016,kaevYadFiz2019}. Here the self-consistent theory of finite Fermi systems (TFFS) \cite{KhSap1982}, based on the effective density functional theory, has been used with  Fayans   functional \cite{Fayans} and PC.  In this sense, one can speak about the second stage of  developing TFFS.  It was shown that in all  the numerous  problems considered, PC contribution is sizable,  of fundamental importance and  necessary for explaining experimental data. The success of these works is explained not only by the use of Fayans functional but also  by a more consistent consideration of  PC effects, in particular, the tadpole effects \cite{KhSap1982,SapTol2016} and, probably, the effect of variation of the effective interaction in the phonon field \cite{voitenkov}.  The important  reason for this success, in the author's opinion, was the use of  consistent quantum many-body formalism, the Green function (GF) method \cite{Migdal},  to be exact. 

 Simultaneously with the above-mentioned GF approach \cite{KhSap1982} and beginning from 1983 \cite{kaev83}, the theory for PDR and GMR was being  developed within GF method and with PC,  in the framework of both non-selfconsistent \cite{kaev83,ts89, revKST} and self-consistent variants      \cite{ts2007,ave2011}. The  difference between \cite{kaev83} and \cite{ts89} consisted in the fact  that in \cite{ts89} the disadvantage of \cite{kaev83} was
eliminated, namely, a special approximate method 
    of chronological decoupling of diagrams (MCDD) (or, in a more modern terminology, time blocking approximation (TBA)) was developed in order to solve the problem of  second order poles in the generalized propagator of the extended RPA propagator in \cite{kaev83}. This disadvantage was not important to explain the properties of M1 resonance \cite{ktZPhys}, but it was important for electrical MGR.
Later on, this method to solve  the second order poles problem was considerably improved \cite{ts2007} so that the approach  obtained the name of  qasiparticle time blocking  approximation (QTBA) for nuclei with pairing. See, for example, \cite{ts2018} for references to the improved TBA approaches
    for magic nuclei. However, the main physical content, i.e. inclusion of PC \textit{only} into the   particle-hole (ph) propagator 
    (in the language of TFFS) has been always preserved  despite the fact that the derivation method  was different and was based on the Bethe-Salpeter equation.   
 In particular, even the known  tadpole effect
 was not considered in the generalized  propagator of  MCDD or TBA,  except for the analytical  attempt in \cite{YadFiz2011}, where this effect was considered within the method of \cite{kaev83}.
  
  Thus, in contrast
to what we have within the non-selfconsistent \cite{Sol89} and  self-consistent \cite{scQPM} quasiparticle–phonon model, where  both energy regions (low-lying and PDR+GMR) are considered on an equal footing, there still exists a gap between GF approaches applied in the region of
the ground state and low-lying excited states \cite{KhSap1982,SapTol2016}, on the one hand, and the approach in (PDR+GMR) energy region  \cite{YadFiz2011, ave2011}, on the other hand. In any case, we are sure  that the possibilities of GF approach have  not been used completely in the microscopic theory of PDR and GMR.

 The goal of this article is to use the  GF approach based on the self-consistent TFFS, in the PDR+GMR  energy region, in order to close this gap and to include  PC consistently, i.e. not only
into the ph-propagator  but also into other quantities  contained in GF apparatus, for example, the effective interaction.
  We will see that such an approach  gives new effects, which can be important for the microscopic description of PDR and GMR. 

In this article, we will consider only  magic nuclei and  complex configurations 1p1h$\otimes$phonon and  discuss only $g^2$ PC corrections, where $g$ is the phonon creation amplitude. As  usual, we use the fact of existence of
the small $g^2$ parameter. Very often we symbolically write  our formulas, the main of which are represented in the form of Feynman diagrams, so the final formulas can be easily obtained. 

\section{Some earlier results with PC}

As it was mentioned in the Introduction, the physical content  of the previous GF approaches  in the PDR and GMR theory was that  $g^2$ PC corrections were included  into ph-propagator, so that the new ph-propagator should correspond  to the diagrams which are  presented  in Fig.1, 
\begin{figure}[h]
\includegraphics[width=0.6\linewidth]{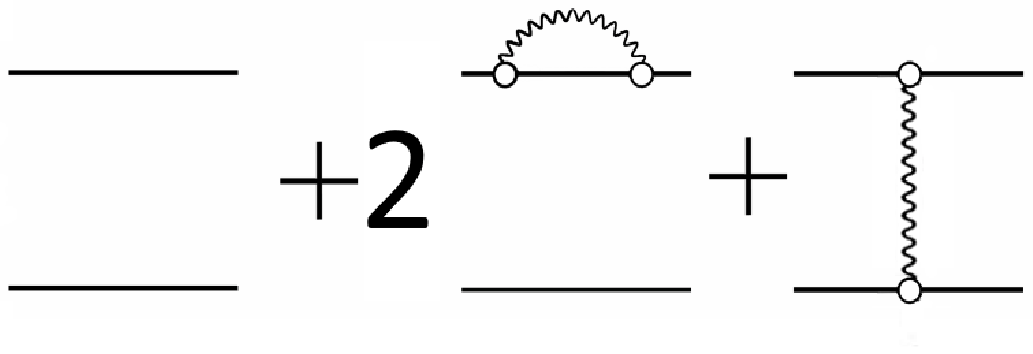}
\caption{Diagrams corresponding to the  simplest ph propagator with PC .}
\label{fig-1}       % Give a unique label
\end{figure}  
where the first diagram means   the RPA case  with the ph propagator 
\bea
A_{12} (\omega) = \int G_1(\eps + \frac{\omega}{2})G_2(\eps - \frac{\omega}{2})d\eps
\label{EQ-propagator}
\eea
Hereinafter number  2 before a graph or a corresponding formula means that there  are two graphs or formulas of a similar type, low indices mean the set of single-particle quantum numbers 1 $\equiv ({n_1,j_1,l_1,m_1})$ , $\omega$ is the energy of the external field $V^0$ and we write everywhere $d\eps$ instead of $\frac{d\eps}{2\imath\pi}$.  Of course, it is always necessary to keep in mind that for  numerical results the method  of MCDD \cite{ts89} or TBA in the case under consideration  should be used in reality to avoid the above-mentioned problem with  second order poles in the ph propagator shown in Fig.1.    Such a    generalized propagator is rather complicated, it   was discussed in details and given in \cite{revKST}. 

In the works, performed within the self-consistent TFFS, the  $g^2$ PC corrections to the mean field, which take into account  the tadpole,  have been used actively.  They can be written in the symbolic form  as
\beq
\delta \Sigma = gDGg + g_{11}D
\label{EQ-delta_sigma}
\eeq
and shown in Fig.2. 
\begin{figure}
\includegraphics[width=0.7\linewidth]{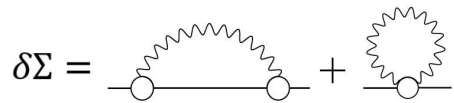}
\caption{The $g^2$ corrections to the mean field. Straight  and wavy lines correspond to GF G and D, circles
with one wavy line stand for the amplitude of phonon production
g, and the circle with two  wavy lines is $g_{11}$}
\label{fig-2}       % Give a unique label
\end{figure}
Here  G and D are the single-particle and  phonon  GF's:
\bea
G_1(\eps) = \frac{1-n_1}{\eps -\eps_1 + \imath\gamma} +  \frac{n_1}{\eps -\eps_1 - \imath\gamma},  \nonumber \\
D_s(\omega) = \frac{1}{\omega - \omega_s + \imath\gamma} - \frac{1}{\omega + \omega_s - \imath\gamma}  
\label{EQ-GF}
\eea
and  g obeys the homogeneous equation \cite{Migdal}
\beq 
g=FAg
\label{Eq-g}
\eeq
where F is the effective interaction , A is the ph propagator, Eq.(\ref{EQ-propagator}). Eq.(\ref{Eq-g}) means  the RPA approach formulated in GF language.  

The amplitude of creation of two  phonons (similar in the tadpole  case) is obtained as the variation of Eq.(\ref{Eq-g}) in  the field of  phonon 2
\bea
g_{12} \equiv \delta^{(1)}g_2 = \delta^{(1)}FAg_1 + F\delta^{(1)}Ag_1 + FAg_{12}
\label{EQ-tadpole}
\eea
The solution of this equation is rather difficult. It has been solved only in the coordinate representation  in the works by A.P. Platonov 
pertaining to other  problems connected with the properties of ground or low-lying collective states \cite{platonov,KhSap1982}. However, it is possible to use a realistic estimation of the two-phonon creation amplitude $g_{11} \equiv \delta^{(1)}g_1$, which is contained in the tadpole term. This estimation has been based on the 
approximation for the quantity $\delta F$ \cite{KhSap1982}:
\beq 
\delta F = (\frc{\delta F}{\delta \rho})\, Ag
\label{EQ-deltaF}
\eeq
If one neglects the small in-volume
term of the amplitude g, a simple expression for $(g_{11})_{ik} = \alpha_L^{2}\partial^{2} U/\partial \textbf{r}_i \partial\textbf{ r}_k$
in the non-pole term $K^{ph} = \int d\omega g_{11}D(\omega)$ can be
obtained \cite{platonov},
where $U$ is the self-consistent mean field and $\alpha_L$ is the deformation parameter of the L-phonon in  the Bohr-Mottelson model .

\section{Equation for new effective field $\tilde V$ with PC corrections}

Our aim is to find $g^2$-corrections to the \textit{ vertex $V$}, which is described by the standard TFFS equation \cite{Migdal} 
 \beq
 V = e_{q}V^0 + FAV
 \label{EQ-V}
 \eeq 
 in order to find, within the GF method,  a new equation for a new vertex, which takes into account all the $g^2$ corrections to $V$. 

We assume that our approach is self-consistent, i.e.  the mean field and  effective interaction $F$  are defined, respectively, 
 as the first and second functional derivations of the effective density functional. In this sense, our approach is a generalization of the self-consistent TFFS with PC corrections to the PDR and GMR energy region.  
 
 Using the general analogy for obtaining Eq.(\ref{EQ-tadpole}) and above-mentioned $g^2$-correction to the mean field, one can write the following 
 for $g^2$-corrections to the   vertex $V$:
  \beq
\tilde V = V + \Delta V(g,  V)
\label{EQ-full_V}
\eeq
and 
%Eq.(9):
 \beq
\Delta V = 2gDG\delta^{(1)}V + \delta^{(2)}VD,
\label{EQ-dV}
\eeq
where the quantities $\delta^{(1)}V$ and $\delta^{(2)}V$ are  the first and second order variations  of the vertex $V$, Eq.(\ref{EQ-V})
in the phonon field.
   They are  shown in Fig.3.  The second term in Eq.(\ref{EQ-dV}),Fig.3, contains  "pure"  $g^2$ corrections, which are, in a sense,  similar to the tadpole corrections in Fig.1,
while the first term  in Eq.(\ref{EQ-dV}),Fig.3, is a mix between the first order correction to the vertex $V$  and the "end" correction of the
   first order in $g$, with  the ends of the first diagram in Fig.3. kept in mind.
   
   \begin{figure}
\includegraphics[width=1\linewidth]{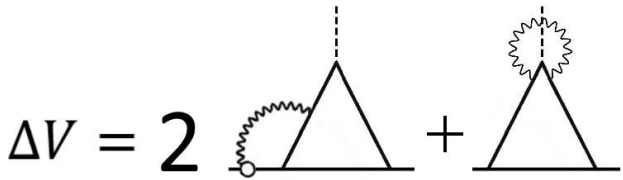}
\caption{The $g^2$ corrections to the vertex V.}
\label{fig-3}       % Give a unique label
\end{figure}
   
   In fact, the corrections $\Delta V$  already appeared in the consideration of quite another problem, namely, the calculations of 
   $g^2$ corrections to  matrix elements $(\varphi_{\lambda_1},V \varphi_{\lambda_2})$ for  static electromagnetic moments of odd nuclei in the ground state \cite{YadFiz2014,JPhysG}.
   This problem was reduced to the analysis of 8 terms with $g^2$ corrections, from which 5 terms contain $\delta^{(2)}\varphi_\lambda$ (the so-called ends corrections) and the other 3 terms are our  corrections $\Delta V$, Eg.(\ref{EQ-dV}), but in the static form,  while in our case we need them at PDR and GMR energies. It turned out 
   that these 5 terms, which depend strongly on single-particle ends $\lambda_1$ and $\lambda_2$, cancel each other  noticeably, so that their algebraic sum gives an improvement  to describe  experimental data under consideration. For some  particular cases, the other 3 corrections were estimated 
   and it was shown that for the problems considered in \cite{YadFiz2014,JPhysG}  the appropriate $g^2$ corrections should not be addressed because they are important not for individual states $\lambda_1 = \lambda_2$,  but for the case when they contain large sums of small contributions.  However, just this latter case of large sums of small contributions is of great interest for us when only  the collective (or "cooperative" \cite{Bracco}) phenomena like PDR and GMR are 
   studied. It is well known that the collectivity of PDR and GMR are due to cogerent sums of many configurations 1p1h + (1p1h$\otimes$phonon) 
   
   Getting back to our problem, let us find the variations $\delta^{(1)}V$ and $\delta^{(2)}V$. But first, we obtain the quantity 
     $\delta ^{(2)}A$ for our case of similar phonons, which is contained in $\delta^{(2)}V$.
    For the general case,   $\delta ^{(2)}A$ is as follows:
\bea
\delta^{(2)}A = \delta^{(2)}G_1G_2 = \delta^{\tilde 2}\delta^{\tilde 1}G_1G_2,
\label{EQ-dA}
\eea
where $\delta^{\tilde 1}$ is variation in the field of phonon 1, and we have introduced  notion $\tilde 1$
for phonon 1, not to confuse it with the single-particle index 1.
Then we have
\bea
\delta^{\tilde 1}G_1G_2 =  \delta^{\tilde 1}G_1 G_2 + G_1 \delta^{\tilde 1}G_2 = \nonumber \\ 
G_1g_{\tilde 1}G_3 G_2 + G_1 G_2g_{\tilde 1}G_3 
\label{EQ-dG_1}
\eea 
\bea
\delta^{\tilde 2}G_1g_{\tilde 1}G_3G_2 = \delta^{\tilde 2}G_1g\delta^{\tilde 1}G_3 G_2 + G_1\delta^{\tilde 2}g_{\tilde 1}G_3 G_2 +  \nonumber \\ 
G_1g_{\tilde 1}\delta^{\tilde 2}G_3G_2 + G_1g_{\tilde 1}G_3\delta^{\tilde 2}G_2
\label{EQ-dG_2}
\eea
and analogically for the second term in Eq.(\ref{EQ-dG_1}). Finally, for the case $\tilde 1 = \tilde 1$, which is of interest in our case  for the variation
$\delta^{(2)}V = \delta^{\tilde 1}\delta^{\tilde 1}V$,  we obtain five terms instead of eight, they  are shown in Fig.4.
\bea
\delta^{\tilde 1}\delta^{\tilde 1}G_1G_2 = 2G_1g_{\tilde 1}G_3g_{\tilde 1}G_4G_2 + \nonumber \\ 
2G_1g_{\tilde 1\tilde 1}G_3G_2 + G_1g_{\tilde 1}G_3G_2g_{\tilde 1}G_4
\label{EQ-dG_end}
\eea

\begin{figure}[h]
\includegraphics[width=0.9\linewidth]{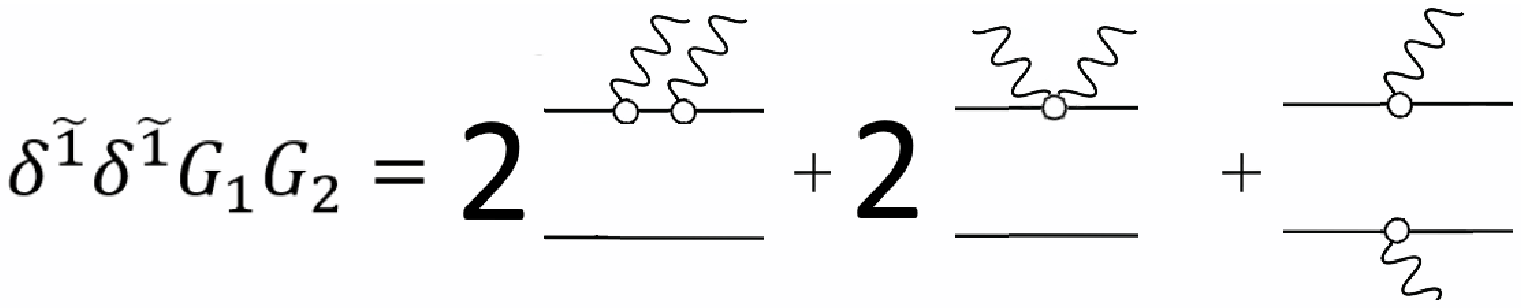}
\caption{ Expression (\ref{EQ-dG_end}) in the diagrammatic representation.}
\label{fig-4}       % Give a unique label
\end{figure}   
   
    The quantities $\delta^{(1)}V$ and $\delta^{(2)}V$ should be obtained by variation  of Eq.(\ref{EQ-V}) in the phonon field:
   
\begin{align}
\delta^{(1)}V =& \delta^{(1)}FAV + F\delta^{(1)}AV + FA\delta^{(1)}V, \nonumber \\ 
\delta^{(2)}V =&  \delta^{(1)}\delta^{(1)}V = F\delta^{(2)}AV + \nonumber \\
&2\delta^{(1)}F\delta^{(1)}AV + 
2\delta^{(1)}FA\delta^{(1)}V 
                +2F\delta^{(1)}A\delta^{(1)}V +  \nonumber \\ 
& \delta^{(2)}FAV + FA\delta^{(2)}V  
\label{EQ-full_dV}
\end{align}
\begin{figure}
\includegraphics[width=1\linewidth]{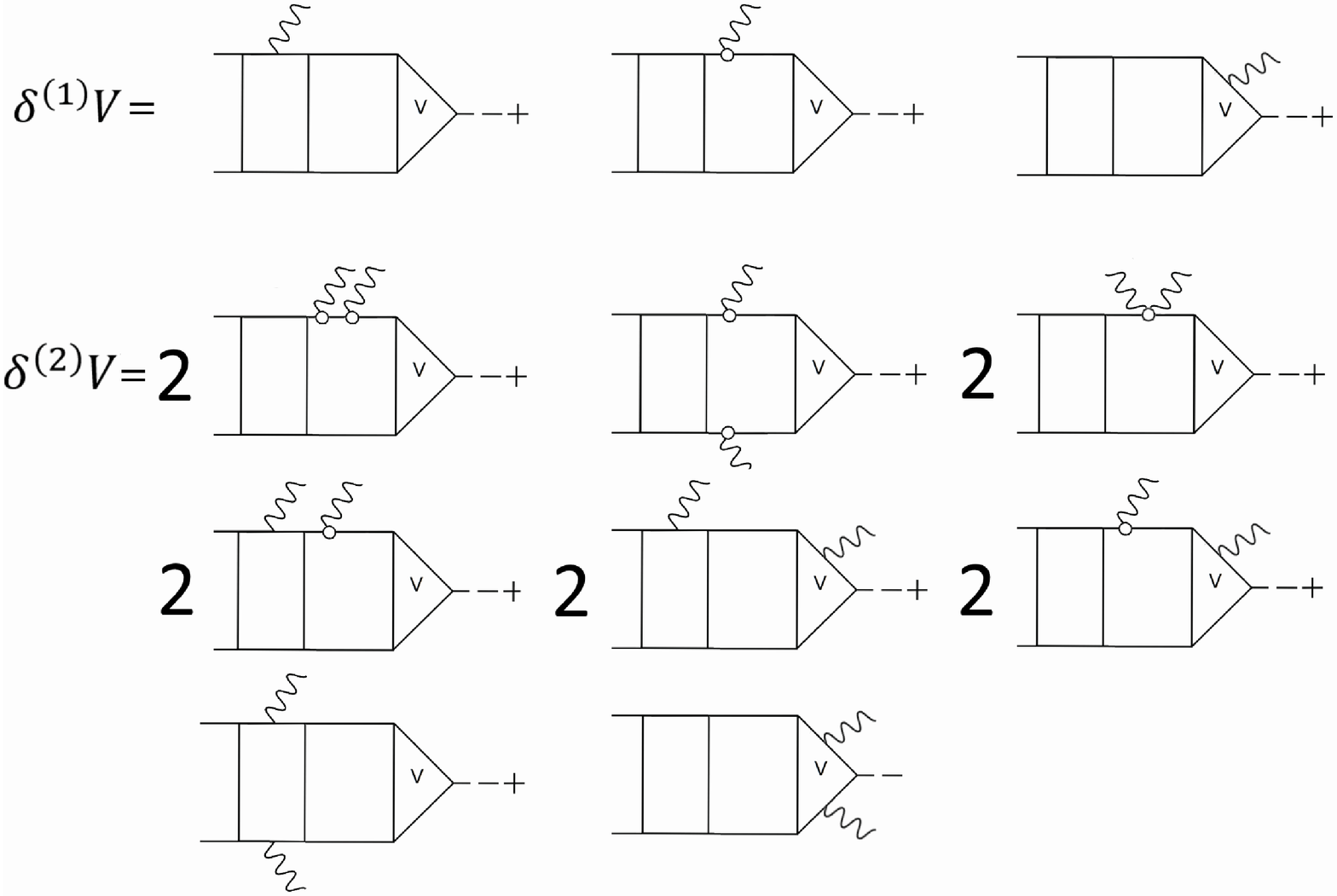}
\caption{First- and second order variations of Eq.(\ref{EQ-V}) in the phonon field.}
\label{fig-5}       % Give a unique label
\end{figure}
They  are shown in Fig.5,
where the term with $\delta ^{(2)}A$ is already shown by the three graphs in the first line for $\delta^{(2)}V$.

 One can see from Eq.(\ref{EQ-full_dV}),Fig.5, that  the quantities $\delta^{(1)}V$ and $\delta^{(2)}V $ obey
a  rather complicated coupled system of integral equations, the first of which has two free terms and the second one has five free terms. All these 
free terms  contain the vertex $V$ or  (in the case of $\delta^{(2)}V $) the quantity $\delta^{(1)}V$. System (\ref{EQ-full_dV}) can be solved if
to use the approximation Eq.(\ref{EQ-deltaF})for $\delta^{(1)}F$  and develop  such an  approximation for $\delta^{(2)}F$ further. 
However, this is a very  
 complicated way and for the better understanding of physical sense, it is better here to use only above-mentioned free terms of
Eq.(\ref{EQ-full_dV}).
 Also, one can transform Eq.(\ref{EQ-full_dV}) in order
to find expressions for $\delta^{(1)}V$ and $\delta^{(2)}V $.  
However, in this case we will obtain a noticeable complication because of appearance of a   two-phonon channel.
 We will not use this way for simplicity and because of our  restriction to only complex 1p1h$\otimes$phonon configurations  mentioned in the Introduction.

 Let us go back to  expression (\ref{EQ-full_V}). One can see that  expression (\ref{EQ-full_V}) is the first iteration of the following equation (if $V$ , Eq.(\ref{EQ-V}) is zero iteration)       
  \beq
\tilde V = V + \Delta V(g, \tilde V).
\label{EQ-full_V+}
\eeq  
Then, after setting of the above-mentioned free terms of Eq.(\ref{EQ-full_dV}) and keeping in mind  Eq.(\ref{EQ-full_V+}),
 we obtain the final equation for the new effective field $\tilde V$:       
\begin{align}
 \tilde V = e_{q}V^0 + FA\tilde V + 2FGgDGgGG\tilde V + FGgGDGgG\tilde V& \nonumber \\ 
            + 2FGg_{\tilde 1\tilde 1}DGG\tilde V + 4gGDFGgGG\tilde V +& \nonumber \\
              + 4FGgDGGFGgGG\tilde V +&                              \nonumber \\ 
            + 2g\delta FDGGG\tilde V + 2\delta FDGgGG\tilde V +&   \nonumber \\ 
                 2\delta FDGGFGgG\tilde V +  2FGgDGG\delta FGG\tilde V +& \nonumber \\
        2\delta FDGG\delta FGG\tilde V + \delta^{(2)}FDGG\tilde V &                                          
 \label{main_EQ}
\end{align}
 It is shown in Fig.6.   
 The lines of Eq. (\ref{main_EQ}) and Fig.6 correspond to each other. Because of the symbolical  form  of Eq.(\ref{main_EQ}), in each of its lines 2 and 3,  we have shown two graphs in Fig.6 instead of the terms with  number  4 in the same   lines of Eq.(\ref{main_EQ}) 
  All the terms in Eq. (\ref{main_EQ}) correspond to the case when only the complex configurations 1p1hx$\otimes$phonon are taken into account in magic nuclei. 
 
 Thus, we have obtained  that all the terms in the first line of Eq.(\ref{main_EQ}),Fig.6, correspond to the previous approach 
 \cite{kaev83,ts89,Tselyev2016}, which  uses the ph-propagator with PC shown in Fig.1. The other lines contain the following difference from this  approach. Namely, these  are contributions  due to: i)   tadpole effect  in the standard  particle-hole propagator (the first term of the second line), ii)  two new induced interactions ( due to  exchange by the usual particle-hole phonon)  in the second particle-hole channel and in the particle-particle channels (the second and third lines), and  iii)  the first and second variations of the effective interaction in the phonon field (the forth, fifth and sixth  lines).
 \begin{figure}[h]
\includegraphics[width=1\linewidth]{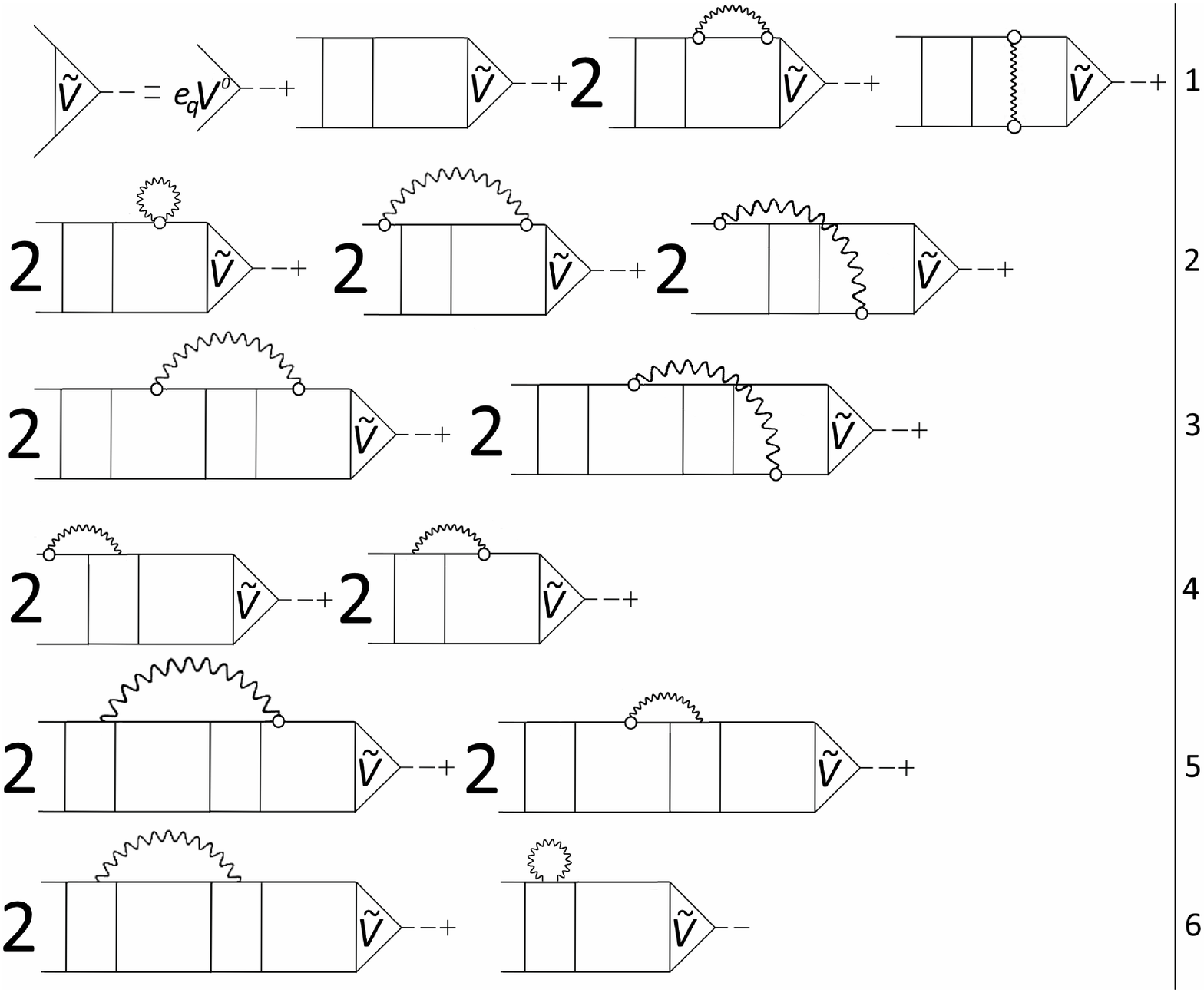}
\caption{Equation (\ref{main_EQ}) for the new effective field $\tilde V$ with PC  in the diagrammatic representation.
For convenience,  numeration  on the right gives numbers of the lines here and in Eq.(\ref{main_EQ})  }
\label{fig-6}       % Give a unique label
\end{figure}

\section{Discussion of the equation for  the new effective field $\tilde V$}

For simplicity, we enumerate the terms  of Eq. (\ref{main_EQ}),Fig.6,  in accordance with their lines as follows: 
 \beq
\tilde V = \tilde V^1 +  \tilde V^{tad} + \tilde V^2 + \tilde V^3 + \tilde V^4 + \tilde V^5 + \tilde V^6
\label{notations}
\eeq 
Here the upper indices mean only the number of the line in  Eq. (\ref{main_EQ}),Fig.6. The separate parts of Eq.(\ref{notations})  may include two or three terms in each line.

 1. The first term, $\tilde V^1$, corresponds to  previous approaches \cite{kaev83,ts89,Tselyev2016}, which  use the ph- propagator with PC shown in Fig.1. For the completeness, we give  the term of $\tilde V^1$, which contains the second part of $\tilde V^1$ (with insertions). The third part (with the old induced interaction) will be discussed in section B. See also Ref.\cite{revKST} for detailed discussions of the TBA approaches.
%\textbf{2 novyx formuly ?!}  
  \beq
\tilde V^{1(2)} = \sum_{3465,s}F_{1263}g_{34}g_{45}\tilde V_{56}I_{3456,s}^{(1)}(\omega)
 \label{EQ-V_1(2)}
 \eeq
 where 
 \begin{align}
&I_{3456,s}^{(1)}(\omega)\!\! =\!\!\!\! \int\!\!\! G_3(\eps)G_5(\eps)G_6(\eps-\omega)\times \nonumber\\
 &G_4(\eps-\omega_1)D_s(\omega_1)  d\eps d\omega_1\nonumber\\
&=\frac{1}{(\eps_{63}+\omega)\eps_{53}}\cdot\Biggl(\frac{n_6(1-n_4)}{\eps_{46}+\omega_s-\omega}-\frac{n_4(1-n_6)}{\eps_{46}-\omega_s-\omega}+\nonumber\\
&\frac{n_4(1-n_3)}{\eps_{43}-\omega_s}-\frac{n_3(1-n_4)}{\eps_{43}+\omega_s}\Biggr)-\nonumber\\
&-\frac{1}{(\eps_{65}+\omega)\eps_{53}}\cdot\Biggl(\frac{n_6(1-n_4)}{\eps_{46}+\omega_s-\omega}-\frac{n_4(1-n_6)}{\eps_{46}-\omega_s-\omega}+\nonumber\\
&\frac{n_4(1-n_5)}{\eps_{45}-\omega_s}-\frac{n_5(1-n_4)}{\eps_{45}+\omega_s}\Biggr)
 \label{EQ-I_3456_1}  
 \end{align}
 Hereinafter   $\eps_{34} = \eps_{\lambda_3} -  \eps_{\lambda_4}, n_3 = n_{\lambda_3}$,
 the low indices mean the  set 1$\equiv ({n_1,j_1,l_1,m_1})$,  and we write down only one  
 term of Eq.(\ref{main_EQ}),Fig.6, which contains   number 2.
 
 Below we discuss  the new terms of Eq.(\ref{main_EQ}), Fig.6.
 
 2. In the second lines of Eq.(\ref{main_EQ}),Fig.6, we obtained the tadpole contribution $\tilde V^{tad}$ and four graphs with the general structure which is similar  to the PC graphs of the first line  in the sense that all of them contain four GFs G, one phonon GF D and two phonon creation amplitudes $g$.
 
 3. In the third  line, there are rather complex graphs with two effective interactions F, six GFs G, one phonon GF D and two phonon creation amplitudes $g$.
 
  All  the 4-th, 5-th and 6-th lines contain the variations of the effective interaction $\delta F$ of the first  and second orders: 
  
 4. In the fourth  line, rather simple graphs with three GFs G , one phonon GF D and one phonon creation amplitude $g$ are present.
  
 5. In the fifth   line, we have rather complex graphs which contain simultaneously the effective interaction F, its variations   $\delta F$
  and one phonon creation amplitude $g$.
 
 6. Finally, in the last,    sixth line,  there are one GF D, two first order variations  of the effective interaction $F$, four GFs G 
 (the first graph) and  the second order variation of  $F$  $\delta^{(2)}F$ with two GFs G and one GF D(the second graph).

 \subsection{The term with tadpole }
    
Let us discuss  the  term in  the second  line of Eq.(\ref{main_EQ}),Fig.6 ($\tilde V^{tad}$  of Eg. (\ref{notations})). One can see, it has 
      a   relatively simple structure:
 
 \bea
 \tilde V^{tad}_{12} = \Sigma F_{1234}\tilde V_{54}K^{ph}_{35}I_{354}(\omega)  
 \label{EQ-V_tad}
 \eea
  \bea
 I_{354}(\omega) = \int G_3(\eps)G_5(\eps)G_4(\eps - \omega)d\eps =\nonumber\\ \frac{n_3-n_4}{(\eps_{34}-\omega)(\eps_{54}+\omega)}+(1-\delta_{35})\frac{n_3-n_5}{\eps_{35}(\eps_{54}+\omega)}
 \label{EQ-I354}
 \eea
 
 In the works of the Kurchatov institute group, it was shown that, as a rule, 
 the quantitative tadpole contributions to characteristics of the ground and low-lying states are rather considerable and have the opposite sign as compared to the pole (usual) diagrams. One can think that  tadpole term $\tilde V^{tad}_{12}$  should give the same effect as compared to
 the insertion terms in $\tilde V^1$ of Eq.(\ref{notations}).

   \subsection{Terms $\tilde V^2$ with new induced interactions}
 
 As one can see from Fig.6, the term $\tilde V^2$ of Eq.(\ref{notations}) contains two new induced interactions caused by the exchange of  our
 phonon  in the second ph-channel and in the particle-particle channels. We call these new terms of
  $\tilde V^2$ as $\tilde V^{ph2}$ and $\tilde V^{pp}$ because the term in $\tilde V^1$
  with the old induced interaction (caused by the exchange of the same  ph-phonon) can be called as  $\tilde V^{ph1}$ and  is already contained in the last term of the first line of Fig.6. All of them are shown in Fig.7: the old one  $\tilde V^{ph1}_{12}$ and two new ones $\tilde V^{ph2}_{12}$ and $\tilde V^{pp}_{12}$. Each of the induced interactions, by our definition,  contains the effective interaction $F$, two single-particle GFs G ,   one phonon GF D , two phonon creation amplitudes $g$ and depends on the single-particle energy variable $\eps_1$. This dependence  corresponds to the time variable of  particle 1 in the $\delta$-function $\delta (\eps_1 - \eps_2 - \omega)$, which  always appears in all final formulas for 
  $\tilde V$.     
  
  \begin{figure}[h]
\includegraphics[width=1\linewidth]{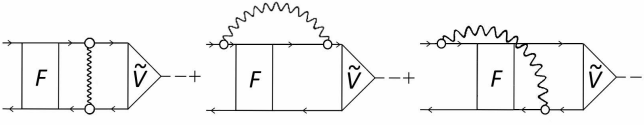}
\caption{Terms of  Eq.(\ref{main_EQ}),Fig.6 with induced interactions:the old one  $\tilde V^{ph1}_{12}$ and new $\tilde V^{ph2}_{12}$ and $\tilde V^{pp}_{12}$.}
\label{fig-7}       % Give a unique label
\end{figure} 

The old term  $\tilde V^{ph1}_{12}$, which is contained in the term $\tilde V^{1}$  of Eq.(\ref{notations}), is as follows:
\beq
\tilde V^{ph1}_{12} =  \sum_{3465,s}F_{1243}g_{35}g_{65}\tilde V_{56}I_{3456,s}^{(2)}(\omega)
 \label{EQ-V_ph1}
 \eeq
 where 
 \begin{align}
&I_{3456,s}^{(2)}(\omega)\!\! =\!\!\!\! \int\!\!\! G_3(\eps)G_6(\eps-\omega)D_s(\omega_1) \times \nonumber\\
 &G_4(\eps-\omega_1)G_5(\eps-\omega_1-\omega) d\eps d\omega_1\nonumber\\
&=\frac{1}{(\eps_{63}+\omega)(\eps_{45}-\omega)}\cdot\Biggl(\frac{n_6(1-n_4)}{\eps_{46}+\omega_s-\omega}-\frac{n_4(1-n_6)}{\eps_{46}-\omega_s-\omega}+\nonumber\\
&\frac{n_5(1-n_3)}{\eps_{35}+\omega_s-\omega}-\frac{n_3(1-n_5)}{\eps_{46}-\omega_s-\omega}+\frac{n_6(1-n_5)}{\eps_{65}-\omega_s}-\nonumber\\
&\frac{n_5(1-n_6)}{\eps_{65}+\omega_s}+\frac{n_4(1-n_3)}{\eps_{43}-\omega_s}-\frac{n_3(1-n_4)}{\eps_{43}+\omega_s}\Biggr)
 \label{EQ-I_3456_2}  
 \end{align}
 We have applied  the simple and  useful formula:
 \bea
 G_{\lambda}(\eps)\cdot G_{\lambda'}(\eps+\omega)=\frac{G_{\lambda}(\eps)-G_{\lambda'}(\eps+\omega)}{\eps_{\lambda}-\eps_{\lambda'}+\omega}
 \label{EQ-G_summ}
 \eea
 
 In the microscopic theory of GMR,  it is well known  that , as a rule, the term  $\tilde V^{ph1}_{12}$  has the  opposite sign as compared to
 insertion terms in $\tilde V^1$ and the maximal  cancellation is for E0 resonances.  In the previous section A, it was said that the quantitative contribution of  tadpole terms has the opposite sign as compared to contributions of pole diagrams. By analogy,  one can think that the relative role of $ \tilde V^{ph1}_{12}$ should be increased as compared to insertion terms in $\tilde V^1$.  Thus, the quantitative final contribution  of our new tadpole terms may be of great interest.       
 
 $\tilde V^{ph1}_{12}$ does not depend on the energy variable, while the new terms  $\tilde V^{ph2}_{12}$ and $\tilde V^{pp}_{12}$ depend on the energy variable $\eps_1$:
 \beq
 \tilde V^2_{12}(\eps_1,\omega) = \tilde V^{ph2}_{12}(\eps_1,\omega) + \tilde V^{pp}_{12}(\eps_1,\omega),
 \label{EQ-V_2}
 \eeq
 
 \bea
 \tilde V^{ph2}_{12}(\eps_1,\omega) = \sum_{3465,s}g_{13}g_{46}F_{3254}\tilde V_{65}I^{(1)}_{3465,s}(\eps_1,\omega),
 \label{EQ-V_ph2}
 \eea
 \bea
 \tilde V^{pp}_{12}(\eps_1,\omega) =  \sum_{3465,s}g_{13}g_{65}F_{3254}\tilde V_{46}I^{(2)}_{3465,s}(\eps_1,\omega).
  \label{EQ-V_pp} 
 \eea
where 
 %Using Eq.(\ref{EQ-G_summ}) we obtain
 \bea
 I^{(1)}_{3465,s}(\eps_1,\omega) =\! \!\int\! I_{465}(\omega_1,\omega)G_3(\eps_1,-\omega_1)D_{s}(\omega_1)d\omega_1
 \label{EQ-I1_3456}
 \eea
 \bea
 I^{(2)}_{3465,s}(\eps_1,\omega)\! =\!\!\! \int\!\!\! I_{465}(\!-\omega,\!-\omega_1)G_3(\eps_1,\!-\omega_1)D_{s}(\omega_1)d\omega_1
  \label{EQ-I2_3456}
 \eea
 \bea
  I_{465}(\omega_1,\omega) = \int G_{6}(\eps)G_{5}(\eps-\omega)G_{4}(\eps-\omega_1)d\eps = \nonumber\\\frac{1}{\varepsilon_{54}+\omega_1-\omega}\cdot\Biggl[ \frac{n_6-n_5}{\varepsilon_{65}+\omega}-\frac{n_6-n_4}{\varepsilon_{64}+\omega_1}\Biggr]
 \label{EQ-I_465}
 \eea

 As one can see from Eqs.(\ref{EQ-V_ph1}),(\ref{EQ-V_ph2}),(\ref{EQ-V_pp}), each of the the quantities $\tilde V^{ph1}, \tilde V^{ph2}, \tilde V^{pp}$ has
 the similar structure: the effective interaction $F$, vertex $\tilde V$, two amplitudes of phonon creation $g$, four single-particle GFs G and one  phonon GF D. At this stage, one can not see a noticeable quantitative difference between them, but, of course, only calculations should clarify  this statement.
 The difference of the similar quantity  $\tilde V^{4(1)}$, Eq.(\ref{EQ-V4_1}) (see below) from the above-mentioned ones is that $\tilde V^{4(1)}$
 contains $\delta F$ instead of $F$ and this difference  may be noticeable.  The  quantities $ \tilde V^{ph2}, \tilde V^{pp}$ and 
 $\tilde V^{4(1)}$ contain an unexpected  effect - they depend on the energy variable  $\eps_1$. Physically, it is not difficult to understand: after the very first interaction of the external field with a nucleus, 
 one of the quasiparticles of the initial ph-pair may create not the next ph-pair, as in the usual RPA, but a phonon with the creation amplitude
 $g$, and after that  another  ph-pair may interact with other ones through effective interaction. In other words, our approach gives more than earlier, additional  possibilities to create the complex configurations 1p1h$\otimes$phonon and this can be clearly
  seen in  Fig.6.         
 
 The  dependence of the effective field (vertex)  on  the energy variable $\eps_1$ arose for the first time. It is caused by   the  first term of Eq.(\ref{EQ-dV}). Note that this term of Eq.(\ref{EQ-dV}) gives in sum eight terms in Fig.6 (connected with $\delta (1)\tilde V$ in Eq.(\ref{EQ-dV})), including three "dangerous" terms with the induced interactions under consideration.
 
 Such a dependence of the vertex  on  the energy variable $\eps_1$ (or time in the time representation) is quite new  and it should be investigated in the future. At present, one can say  the following. Our vertex $\tilde V$ is , by definition, a change of FG G in the weak external field $V^0$ and the $\eps_1$ dependence arose from the account  for PC for the vertex. There is an analogy with $\eps$-dependence and   PC role in the optical potential  
 $V_{opt}$, which is, in fact, mass operator $\Sigma(\eps)$, for example, see  article \cite{Giai}. By analogy, one can speak about
  $Re\tilde V$ and  $Im\tilde V$. Then   $Im\tilde V$ corresponds to a change of an absorption in the external field, i.e. to the quantity  $ImV_{opt}$. As the  microscopic calculations \cite{Giai} of PC contribution to  $V_{opt}$ showed, 
 %  PC contribution to the   $ReV_{opt}$ gives only a small part to the potential of the shell model, for example, to the Woods-Saxon potential,
   the addition  of PC to  $ImV_{opt}$ turned out too insignificant  to explain the experiment.
 Maybe, one can hope that such  analogy may justify the  suggestion that in the calculations for the beginning one can use the approximation 
  $\eps_1 = \eps_{\lambda_1}$, where $\eps_{\lambda_1}$ is the single-particle energy, i.e. to take $\eps_1$ on the mass surface.

  \subsection{Terms in the third line of Eq.(\ref{main_EQ}),Fig.6 }

 The next terms  of the third line of  Eq.(\ref{main_EQ}),Fig.6 can be expressed through   $\tilde V^{ph2}_{12}$ and $\tilde V^{pp}_{12}$:
 \begin{align}
 \tilde V^{3}_{78} = &\sum_{21}F_{7821}\int G_1(\eps)G_2(\eps - \omega)(\tilde V^{ph2}_{12} + \tilde V^{pp}_{12})(\eps,\omega)d\eps
  \nonumber\\ = 
 &\sum_{213465,s} F_{7821}g_{13}g_{46}F_{3254}V_{65}I^{ph2}_{213465,s} + \nonumber\\
&\sum_{213465,s} F_{7821}g_{13}g_{65}F_{3254}\tilde V_{46}I^{pp}_{213465,s}
 \label{EQ-V_3} 
 \end{align}
 where
 \bea
 I^{ph2}_{213465,s} = \int G_1(\eps)G_2(\eps - \omega) I^{(1)}_{3465,s}(\eps,\omega)d\eps
  \label{EQ-I_ph2}
 \eea
  \bea
 I^{pp}_{213465,s} = \int G_1(\eps)G_2(\eps - \omega) I^{(2)}_{3465,s}(\eps,\omega)d\eps
  \label{EQ-I_pp}
 \eea
 with $I^{(1)}_{3465,s}(\eps,\omega)$ and $I^{(2)}_{3465,s}(\eps,\omega)$ from Eq.(\ref{EQ-I1_3456}) and Eq.(\ref{EQ-I2_3456})

 \subsection{Terms containing $\delta F$}
 \subsubsection{Terms in the fourth line of Eq.(\ref{main_EQ}),Fig.6 }
 
 The first term  $\tilde V^{4(1)}$ of $\tilde V^4$:
 
  \bea
 \tilde V^{4(1)}_{12}(\eps) = \sum_{345,s}g_{13}\delta F_{3254,s}V_{45}I^{(1)}_{345,s}(\eps,\omega)
 \label{EQ-V4_1}
 \eea
 \bea
 I^{(1)}_{345,s}(\eps ,\omega)= A_{45}(\omega)I_{3,s}(\eps), \nonumber \\ 
% A_{45}(\omega)= \int G_4(\eps_4)G_5(\eps_4-\omega)d \eps_4=\frac{n_4-n_5}{\eps_{45}-\omega }, \nonumber \\
 I_{3,s}(\eps) = \int  G_3(\eps-\omega_1)D_s(\omega_1) d\omega_1 = \nonumber\\
  \frac{n_3}{\eps+\omega_s-\eps_3-i\gamma}+\frac{1-n_3}{\eps-\omega_s-\eps_3+i\gamma}
 \label{EQ-I_3s}
 \eea
 Again, we get here the  dependence on the energy variable $\eps$ caused by the first term of Eq.(\ref{EQ-dV}), see the discussion in section B. 
 
 The second term $\tilde V^{4(2)}$  is as follows
 
  \bea
 \tilde V^{4(2)}_{12} = \sum_{345,s}\delta F_{1254,s}g_{34}V_{45}I^{(2)}_{345,s}( \omega)
 \label{EQ-V4_2}
 \eea
\begin{align}
 &I^{(2)}_{345,s}( \omega)=
 \int G_3(\eps-\omega_1)G_4(\eps)G_5(\eps-\omega)D_s(\omega_1) d\eps d\omega_1=\nonumber\\& \!=\frac{1}{\eps_{45}+\omega}\!\!\int\!\! \biggl( G_4(\eps)-G_5(\eps)\biggr) \!\!\!\times\!\!   
 G_3(\eps-\omega_1)D_s(\omega_1) d\eps d\omega_1  = \nonumber\\
 &\frac{1}{\eps_{45}+\omega}\int \biggl( G_4(\eps)-G_5(\eps)\biggr)I_{3s}(\eps)d\eps 
 \label{EQ-I2_345_1}
\end{align}
 with $I_{3s}(\eps)$ from Eq.(\ref{EQ-I_3s})
 
 Thus, for  $I^{(2)}_{345,s}$ we have:
\bea
 I^{(2)}_{345,s}( \omega)=\frac{1}{\eps_{45}+\omega}\Biggl[\frac{n_{4}(1-n_3)}{\eps_{43}-\omega_s}-\frac{(1-n_{4}) n_3 }{\eps_{43}+\omega_s}+  \nonumber\\
 \frac{n_{5}(1-n_3)}{\eps_{53}-\omega_s+\omega}-\frac{(1-n_{5}) n_3 }{\eps_{53}+\omega_s+\omega}\Biggr]
 \label{EQ-I2_345_2}
 \eea
 %here $\eps_{35} = \eps_{\lambda_5} - \eps_{\lambda_3}$.

   \subsubsection{Terms in the fifth line of Eq.(\ref{main_EQ}),Fig.6 }
   These terms contain $\delta F$ and  $F$
      \bea
 \tilde V^{5(1)}_{12}(\omega) = \sum_{34657,s}\delta F_{1234,s}F_{3457}g_{76}V_{65}I_{34765,s} (\omega)
 \label{EQ-V5_1}
 \eea
 \bea
 I_{34765,s}( \omega) = \int A_{34}(\omega - \omega_1)I_{765}(\omega , \omega_1)D_s(\omega_1)d\omega_1
  \label{EQ-I_34765}
 \eea
 with $I_{765}(\omega , \omega_1)$ from Eq.(\ref{EQ-I_465})
 
 The second term of $\tilde V^{5(2)}$ is:
 
 \bea
 \tilde V^{5(2)}_{12}( \omega)\!\!=\!\! \sum_{3456,s}\!\! F_{1234}\!\int\! G_3(\eps)G_4(\eps - \omega)V_{34}^{4(1)}(\eps - \omega)d\eps
 %\delta F_{3654}V_{45}I_{3456,s}^2(\eps_1 \omega)
 \label{EQ-V5_2}
 \eea
 with  $V_{34}^{4(1)}(\eps - \omega)$ from Eq.(\ref{EQ-V4_1})
 
 \subsubsection{The first term in the sixth line of Eq.(\ref{main_EQ}),Fig.6}
  
 Here we have
 \bea
 \tilde V^{6(1)}_{12}(\omega) = \sum_{3456,s}\delta F_{1243,s} \delta F_{3465,s}V_{56}I_{3456,s}( \omega), 
 \nonumber\\
 I_{3456,s}(\omega) =A_{56}(\omega)\cdot\int A_{34}(\omega,-\omega_1) D_{s}(\omega_1)d\omega_1 
 \label{EQ-V6_1}
 \eea
  \subsection{Term containing $\delta^{(2)}F$}
The second term of $\tilde V^6$:
 \bea
 \tilde V^{6(2)}_{12}( \omega) = \sum_{54 }\delta^{(2)} F_{3254}D V_{45}A_{45}(\omega ),
 \label{EQ-V6_2}
 \eea
 where $\delta^{(2)} F_{3254}$ should be obtained with a generalization of  Eq.(\ref{EQ-deltaF}).

 \section{Energies and probabilities of transitions}
 
Below we describe, in a short form,  a general scheme for calculations of the observable characteristics
within our approach. It may be also useful  for calculations and  analysis  of selected  terms  of Eq.(\ref{main_EQ}),Fig.6.    
In order to see more clearly the main features of the method, Eq.(\ref{main_EQ}),Fig.6 can be written in the following form:
  \bea
\tilde V =  e_{q}V^0 + FA^1\tilde V + F(A^{tad} + A^2 + A^3)\tilde V +  \nonumber\\
 \delta F(A^4 + A^5)\tilde V + (\delta F)^{2}A^{6(1)}\tilde V + \delta^{(2)}FA\tilde V
\label{Eq-other}
\eea 
with the obvious notations for the propagators $A^i$, for which  the indices $i$ correspond to the numbers of lines in
Eq.(\ref{main_EQ}),Fig.6 and parts $V^i$ in Eq.(\ref{notations}). Here $A^1$ corresponds to the previous approach \cite{kaev83,ts89} and (without pairing) \cite{ts2007},
which includes the RPA part and PC part with two graphs with insertions and induced interaction $F^{ph1}$ shown in Fig.1 Propagators $A^i$ can be easily 
obtained:  $A^1$ from Eqs.(\ref{EQ-V_1(2)}),(\ref{EQ-V_ph1}), $A^{tad}$ from Eq.(\ref{EQ-V_tad}), $A^2$ from Eqs.(\ref{EQ-V_ph2}),(\ref{EQ-V_pp}), $A^3$ from Eq.(\ref{EQ-V_3}),$A^4$ from Eqs.(\ref{EQ-V4_1}),(\ref{EQ-V4_2}),
$A^5$ from Eqs.(\ref{EQ-V5_1}),(\ref{EQ-V5_2}), $A^{6(1)}$ from Eq.(\ref{EQ-V6_1}). The quantity $A^5$ contains  the effective interaction $F $.

To get formulas for energies and probabilities of transitions between the ground and excited states, we  generalise  the method 
 of standard TFFS \cite{Migdal}. In the pole under consideration $\omega = \omega_n$, the vertex $\tilde V$ has a form
  \beq
\tilde V = \frac{\chi^n}{\omega - \omega_n} + \tilde V^{R} ,
 \label{EQ-V_migdal}
 \eeq
 where $\omega_n$ is the energy of the excited state under consideration,  $\chi^n$ is the residue in this pole and  $\chi^{R}$ is a regular part of $\tilde V$. Then the transition energies should be found from the homogeneous equation
 \bea
\chi^n  =   FA^1\chi^n  + F(A^{tad} + A^2 + A^3)\chi^n  +  \nonumber\\
 \delta F(A^4 + A^5)\chi^n  + (\delta F)^{2}A^{6(1)}\chi^n  + \delta^{(2)}FA\chi^n 
\label{energies}
\eea 
 
It is convenient to rewrite Eq.(\ref{Eq-other}) in a more compact form:
 \beq
\tilde V = e_{q}V^0 + F\overline A\tilde V .
 \label{EQ-V_new}
 \eeq
 where $\overline A $ has  a  complicated form and can be obtained from Eq.(\ref{Eq-other}). It depends on four single-particle indices , can contain the effective interaction $F$, $\delta F$ and 
 different sums. Then, the transition energies should be found from the homogeneous equation:
   \beq
\chi^n = F\overline A^{n}\chi^n ,
 \label{EQ-chi_new}
 \eeq
 where $\overline A^{n} = \overline A(\omega = \omega _n)$

In order to find a normalization condition for the residue $\chi^n$, first, we derive the equation for $\tilde V^{R}$:
\beq
 \tilde V^{R} = e_{q}V^0 + F\overline A^{n}\tilde V^{R} + F\frac{d\overline A^{n}}{d\omega}\chi^n
 \label{EQ-V^R}
 \eeq
 and multiply it on the left by $\chi^n\overline A^{n}$. As a result, we obtain the normalization condition for the residue $\chi^n$:
    \beq
\chi^n\overline A^{n}e_{q}V^0 = -\chi^n\frac{d\overline A^{n}}{d\omega}\chi^n.
 \label{EQ-norma}
 \eeq
 
   Let us obtain  the  probabilities of transitions between the ground  and excited states. This quantity is defined as usual:
   \beq
W_{0n} = 2\pi\mid M_{0n}\mid^2\delta (\omega - \omega_n)
 \label{EQ-W_0}
 \eeq 
 where the matrix element should be found from the polarizability   operator, which contains the change of the density matrix $\rho$ in the external field for the transition from the ground to  excited state $n$:
   \beq
\Pi(\omega) = \sum_{12}V^{0}_{12}\rho_{21}.
 \label{EQ-polar}
 \eeq
 Further, it is quite natural to define our density matrix $\rho$ in the following  form:
  \beq
\rho = \overline A\tilde V
 \label{EQ-rho}
 \eeq
 One can show  that this definition is the same, within our $g^2$ approximation,
as the usual definition $\rho = \int GG\tilde Vd\epsilon$. So, from Eq.(\ref{EQ-V_new}) we have the equation for $\rho$:
 \beq
\rho = \overline A e_{q}V^0 + \overline A F\rho
 \label{EQ-rho_new}
 \eeq
 
 The transition probabilities should be obtained as the residue of the polarizability operator 
 \beq
W_{0n} = 2Im(e_{q}V^0\rho)^n
 \label{EQ-W_rho}
 \eeq
 with the residue $\eta^n$ of $\rho$ in the pole $\omega_n$
 \beq
\rho = \frac{\eta^n}{\omega - \omega_n} + \rho^R,
 \label{EQ-rho_inpole}
 \eeq
 so that from  Eq.(\ref{EQ-V_migdal}) we have 
 \beq
\rho = \overline A (\frac{\chi^n}{\omega - \omega_n} + V^R)
 \label{EQ-rho_end}
 \eeq
 and for the residue $\eta^n$ we find 
   \beq
\eta^n = \overline A\chi^n
 \label{EQ-eta^n}
 \eeq
 Finally, we obtain :
 \beq
W_{0n} =  (e_{q}V^0\overline A\chi^n)
 \label{EQ-W_new}
 \eeq
 or, accounting for the normalization condition Eq.(\ref{EQ-norma}): 
 \beq
W_{0n} = \frac{(e_{q}V^0\overline A\chi^n)^2}{(-\chi^n\frac{d\overline A}{d\omega}\chi^n)}
 \label{EQ-W_final}
 \eeq
 
 For the  energy regions, for which   it is not possible to study the individual eigenenergies of the states,  it is necessary to have an envelope of GMR under consideration. Then 
 it makes  sense to use the smearing parameter $\Delta$, which greatly reduces numerical difficulties of the calculations, if, of course,  a peaked function has a width substantially
larger than the energy averaging interval,  see \cite{revKST}. In this case one uses the strength function  
 \beq
S(\omega, \Delta) = \frac{dB(EL)}{d\omega} = -\frac{1}{\pi}Im\sum_{12}e_{q}V^{0}_{12}\rho_{21}(\omega + \imath\Delta), 
 \label{EQ-PSF}
 \eeq
 from which one can easily  obtain the transition probabilities and  energy-weighed sum rule, summed over an energy interval.  
 
 The formulas obtained in this section are very general  and can be also used, if necessary, to study  any selected part of Eq.(\ref{notations})
 or  Eq.(\ref{EQ-V_new}).    

\section{Conclusion}
In this work, the formalism of many-body nuclear self-consistent theory,  the quantum GF method, to be exact, has been applied for PDR and GMR in magic nuclei to take PC effects and complex 1p1h$\otimes$phonon configurations  into account. The main physical difference from the previous approaches is that, 
due to  consistent inclusion of PC effects, our approach gives much  more than earlier additional possibilities to create the complex configurations 1p1h$\otimes$phonon and that can be cleary seen from  Fig.6. It is necessary to note that within the method under consideration, one can take into account the single-particle spectrum \cite{revKST} and, which is more important, all the new (as compared to the case of RPA ) numerous  ground state correlations (GSC), including three-quasiparticle GSC , for example, in $\tilde V^{tad}$ of Eq.(\ref{notations}),  and more complex GSC. The effects of new three- and four-quasiparticle GSC were disscused in  \cite{voitenkov,picma} for the case of ground and low-lying states and, as it turned out, they are very considerable.

A new equation for the effective field with PC and quite  new PC contributions to the effective field, which  are of interest in the energy regions of PDR and GMR, have been obtained. 
 These contributions are: i)the tadpole effect in the standard ph-propagator, ii)two new induced interactions due to phonon exchange in the second ph-channel (in addition to the old induced interaction in the first ph-channel ) and the induced interactions in the pp- and hh-channels,
iii) the effects of the first and second variations of the effective interaction in the phonon field.\footnote{Generally speaking, for  more exact derivation of the equation for the effective field with PC, it is necessary 
 to refuse from the restriction to  the complex 1p1h$\otimes$phonon configurations only. In this case, the general structure of the new equation will be similar to Eq.(\ref{main_EQ}) and, what is important, all the new above-mentioned ingredients  of our approach will remain the same.} 

 Thus, we have extended the above-mentioned self-consistent approach \cite{KhSap1982} to the 
energy region of PDR and GMR in order to describe \textit{on the equal footing} both the ground states and all the region of nuclear excitations up to GMR energies (30-35 MeV). 
This extension does not concern  only the  PDR and MGR energy region, but also generalizes the TFFS formalism.
In this sense, one can speak about the  beginning of the third stage of developing TFFS.
   As it was mentioned in the Introduction, and we would like to stress it again that, if necessary, the prescription of MCDD, or TBA  \cite{ts89}
should be used in calculations.
  Unfortunately, at present there is no other method to solve the second order poles problem within the formalism considered. 

We have considered, rather schematically, all the  terms in the new equation for the effective field $\tilde V$ Eq.(\ref{main_EQ}),Fig.6.
Strictly speaking, at the present stage it is not possible to speak  about numerical  contributions of the separate terms in view of the fact
that the  approach  is applicable for many
physical cases (the multipole orders, energies of transitions, of excitations etc.). 
%In other
%words, we have many new physical situations
%a big variety? of multipoles ? ,energies and nuclei for which our approach can be applied .?
 One can think, however, that the effects of the tadpole 
in the part $\tilde V^{tad}$ and of the new induced interactions in the parts $\tilde V^2$ and maybe (with $\delta F$) in $\tilde V^4$ can give something quite new
both in qualitative and quantitative sense, while other terms with $\delta F$ and  the term $\delta^{(2)}F$ can  be not so important. 

Probably, our new effects are of the highest interest for calculations of  fine structures of GMR and, especially, PDR and other pygmy-resonances. As it was said in the Introduction, 
there are new physical phenomena in the PDR energy region  and new experimental  methods for fine structure studies, but there is no  reasonable self-consistent explanation of the PDR fine structure even for $^{208}$Pb. However, the problem of the self-consistent explanation of GMR is topical too, first of all, for M1 resonances, where the problem with the second order poles are not so important \cite{ktZPhys}. All these problems will be discussed by us in the  future.

\section{ ACKNOWLEDGMENTS}

We are grateful to V.A. Khodel ,S.V.Tolokonnikov and V.I. Tselayev  for  useful discussions 
 and to S.S. Pankratov for discussions of calculation 
problems. 
The reported study was funded by RFBR, project no.19-31-90186
 and supported by the Russian Science Foundation, project no.16-12-10155.

%\newpage


\begin{thebibliography}{}
\bibitem{Paar} N. Paar, D. Vretenar, E. Khan, G. Colo, Rep. Prog. Phys. \textbf{70}, 691 (2007).
\bibitem{Bracco}
A. Bracco, E.G. Lanza, and A. Tamii, Prog. Part. Nucl.Phys. \textbf{106}, 360 (2019).
\bibitem{kaevYadFiz2019}
S.P. Kamerdzhiev, O.I. Achakovskiy, S.V. Tolokonnikov, M.I. Shitov, Phys. At. Nucl. \textbf{82}, 366 (2019)

\bibitem{revKST} S. Kamerdzhiev, J. Speth, G. Tertychny, Phys. Rep. \textbf{393}, 1 (2004).



\bibitem{ave2011} A. Avdeenkov, S. Goriely, S. Kamerdzhiev, S. Krewald, Phys. Rev. C \textbf{83}, 064316 (2011).

%6
\bibitem{Goriely}
 S. Goriely, E. Khan, V. Samyn, Nucl. Phys. A \textbf{739}, 331 (2004).
\bibitem{YadFiz2011}
S.P. Kamerdzhiev, A.V. Avdeenkov, D.A.Voitenkov, Phys. At. Nucl.\textbf{ 74},  1478 (2011).

\bibitem{Tselyev2016}
V. Tselyaev, N. Lyutorovich, J. Speth, S. Krewald,
and P.-G. Reinhard, Phys. Rev. C \textbf{94}, 034306 (2016).
\bibitem{Lyutor2018}
N. A. Lyutorovich, V. I. Tselyaev, O. I. Achakovskiy,
and S. P. Kamerdzhiev, JETP Lett. \textbf{107}, 659 (2018).
\bibitem{Nester}
A. Repko, V.O. Nesterenko, J. Kvasil, and P.-G. Reinhard, arXiv:1903.01348 [nucl-th] (2019).
\bibitem{Cosel}
A. Tamii, I. Poltoratska, P. von Neumann-Cosel, Y. Fujita, T. Adachi, C. A. Bertulani, J. Carter, M. Dozono, H. Fujita, K. Fujita, K. Hatanaka, D. Ishikawa, M. Itoh, T. Kawabata, Y. Kalmykov, A. M. Krumbholz, E. Litvinova, H. Matsubara, K. Nakanishi, R. Neveling, H. Okamura, H. J. Ong, B. Özel-Tashenov, V. Yu. Ponomarev, A. Richter, B. Rubio, H. Sakaguchi, Y. Sakemi, Y. Sasamoto, Y. Shimbara, Y. Shimizu, F. D. Smit, T. Suzuki, Y. Tameshige, J. Wambach, R. Yamada, M. Yosoi, and J. Zenihiro, Phys. Rev. Let. \textbf{107}, 062502 (2011).
\bibitem{Larsen}
A. C. Larsen, J. E. Midtbø, M. Guttormsen, T. Renstrøm, S. N. Liddick, A. Spyrou, S. Karampagia, B. A. Brown, O. Achakovskiy, S. Kamerdzhiev, D. L. Bleuel, A. Couture, L. Crespo Campo, B. P. Crider, A. C. Dombos, R. Lewis, S. Mosby, F. Naqvi, G. Perdikakis, C. J. Prokop, S. J. Quinn, and S. Siem, Phys. Rev. C \textbf{97}, 054329 (2018).

%13
\bibitem{SapTol2016}
E. E. Saperstein and S. V. Tolokonnikov, Yad.Fiz.\textbf{ 79}, 703 (2016) [Phys. At.Nucl. 79, 1030 (2016)].
%14
\bibitem{KhSap1982}
V. A. Khodel and E. E. Saperstein, Phys. Rep. \textbf{92},183 (1982).

%15
\bibitem{Fayans} A.V. Smirnov, S.V. Tolokonnikov, S.A. Fayans
 Sov. J. Nucl. Phys. \textbf{48}, 995(1988).
\bibitem{voitenkov}
D.Voitenkov, S. Kamerdzhiev, S. Krewald, E.E. Saperstein, S.V. Tolokonnikov, Phys. Rev. C \textbf{85}, 054319 (2012).
\bibitem{Migdal}
A. B.Migdal, Theory of Finite Fermi Systems and
Applications to Atomic Nuclei (Nauka, Moscow,
1965; Intersci., New York, 1967).

\bibitem{kaev83}
S.P. Kamerdzhiev, Yad. Fiz. \textbf{38}, 316 (1983) [Sov. J. Nucl. Phys. \textbf{38},188 (1989) ].
\bibitem{ts89}
V. I. Tselyaev, Yad. Fiz. \textbf{50}, 1252 (1989) [Sov. J. Nucl. Phys. \textbf{50}, 780 (1989) ].

%20
\bibitem{ts2007} V. Tselyaev, 
Phys. Rev. C \textbf{75}, 024306 (2007)

\bibitem{ktZPhys}
S. P. Kamerdzhiev and V.N. Tkachev, Z. Phys.\textbf{A334}, 19 (1989).

\bibitem{kaev2014}
 S.P. Kamerdzhiev, A.V. Avdeenkov, O.I. Achakovskiy, Phys. Atom. Nucl. \textbf{77}, 1303 (2014).

%23

\bibitem{ts2018}
V. Tselayev, N. Lyutorovich, J. Speth,  P.-G.Reinhard, 
 Phys.Rev. C \textbf{97}, 044308 (2018)
\bibitem{Sol89} V.G. Soloviev, {\it Theory of atomic nuclei: quasi-particles and phonons} (Institute of physics, Bristol and Philadelphia, USA, 1992).
\bibitem{scQPM}
Nguyen Van Giai, Ch. Stoyanov, and V. V. Voronov,
Phys. Rev. C\textbf{ 57}, 1204 (1998). 
\bibitem{platonov}
 V. A. Khodel, A.P. Platonov, E.E. Saperstein, J. Phys. G: Nucl. Phys.\textbf{ 6}  1199 (1980).
 \bibitem{YadFiz2014}
 E. E. Saperstein, O. I.Achakovskiy, S. P. Kamerdzhiev,
S. Krewald, J.Speth, and S. V. Tolokonnikov,
 Physics of At. Nuclei,  \textbf{77},  1033 (2014). 
 \bibitem{JPhysG}
 E.E. Saperstein, S.P. Kamerdzhiev, D. S. Krepish, S. V. Tolokonnikov and D. Voitenkov,
 J. Phys. G: Nucl. Part. Phys. \textbf{44},  065104 (2017).
 \bibitem{Giai}
 V. Bernard and Nguyen Van Giai, Nucl. Phys. \textbf{A327}  397, (1979)
 \bibitem{picma}
S.P. Kamerdzhiev, D.A.Voitenkov, E.E. Saperstain, S.V. Tolokonnikov, M.I. Shitov,
 JETP Lett, \textbf{106}, No.3, 139 (2017).


\end{thebibliography}
\end{document}